\documentclass[doublecol,linenumbers]{epl2} 
\usepackage{graphicx,color}  
\usepackage[english]{babel} 
\usepackage{epsfig,dsfont,amssymb,amsmath,amsthm,amsfonts,amsbsy,mathrsfs}
\usepackage{xcolor}
\newcommand{\be}{\begin{equation}}
\newcommand{\ee}{\end{equation}}
\newcommand{\ba}{\begin{eqnarray}}
\newcommand{\ea}{\end{eqnarray}}
\newcommand{\bml}{\begin{multline}}
\newcommand{\eml}{\end{multline}}

\newcommand{\mm}{\mathcal{M}}

\usepackage{hyperref}

\definecolor{red}{rgb}{1,0,0}
\definecolor{blue}{rgb}{0,0,1}
\definecolor{black}{rgb}{0,0,0}

\usepackage{amsmath}

\title{On Variational Arguments for Vibrational Modes near Jamming}

\author{Le Yan\thanks{lyan@kitp.ucsb.edu}\inst{1} \and Eric DeGiuli\inst{2} \and Matthieu Wyart\thanks{matthieu.wyart@epfl.ch}\inst{2}}

\institute{                    
  \inst{1} Kavli Institute for Theoretical Physics, University of California - Santa Barbara, CA 93106, USA\\
  \inst{2} Institute of Theoretical Physics, \'Ecole Polytechnique F\'ed\'erale de Lausanne - CH-1015 Lausanne, Switzerland
}

\abstract{
Amorphous solids tend to present an abundance of soft elastic modes, which diminish their transport properties,
generate heterogeneities in their elastic response, and affect non-linear processes like thermal activation of plasticity. 
This is especially true in packings of particles near their jamming transition, for which effective medium theory and variational arguments can both predict the density of vibrational modes. However, recent numerics support that one  hypothesis of the variational argument does not hold.  We provide a novel variational argument which overcomes this problem, and correctly predicts the scaling properties of soft modes near the jamming transition. Soft modes are shown to be related to the response to a local strain in more connected networks, and to be characterized by a volume $1/\delta z$, where $\delta z$ is the excess coordination above the Maxwell threshold. These predictions are verified numerically. 
}

\begin{document}

\maketitle

\section{Introduction}
Most amorphous solids present an excess of vibrational modes over the Debye model,  the ``boson peak'' \cite{Phillips81}.  Such soft elastic modes strongly affect transport  \cite{Baldi10,Xu09} as well as linear elastic response and its fluctuations \cite{Leonforte06,Lerner14,Karimi15,Ellenbroek06}. They also contribute to non-linear properties such as thermal activation \cite{Brito07,Hocky12} and plasticity \cite{Manning11}. In some cases, they even control the structure of the material, including in soft elastic particles near jamming \cite{Wyart05a} and  dense colloidal glasses \cite{ Brito09,DeGiuli14b} where the structure is marginally stable: soft elastic modes different from plane waves are present down to zero frequency. 
Beyond  amorphous solids,  soft elastic modes are also central to designed materials with controlled \cite{Florijn14} and sometimes topological properties \cite{Kane14,Paulose15}. 

\begin{figure}[!ht]
\centering
\includegraphics[width=1\columnwidth]{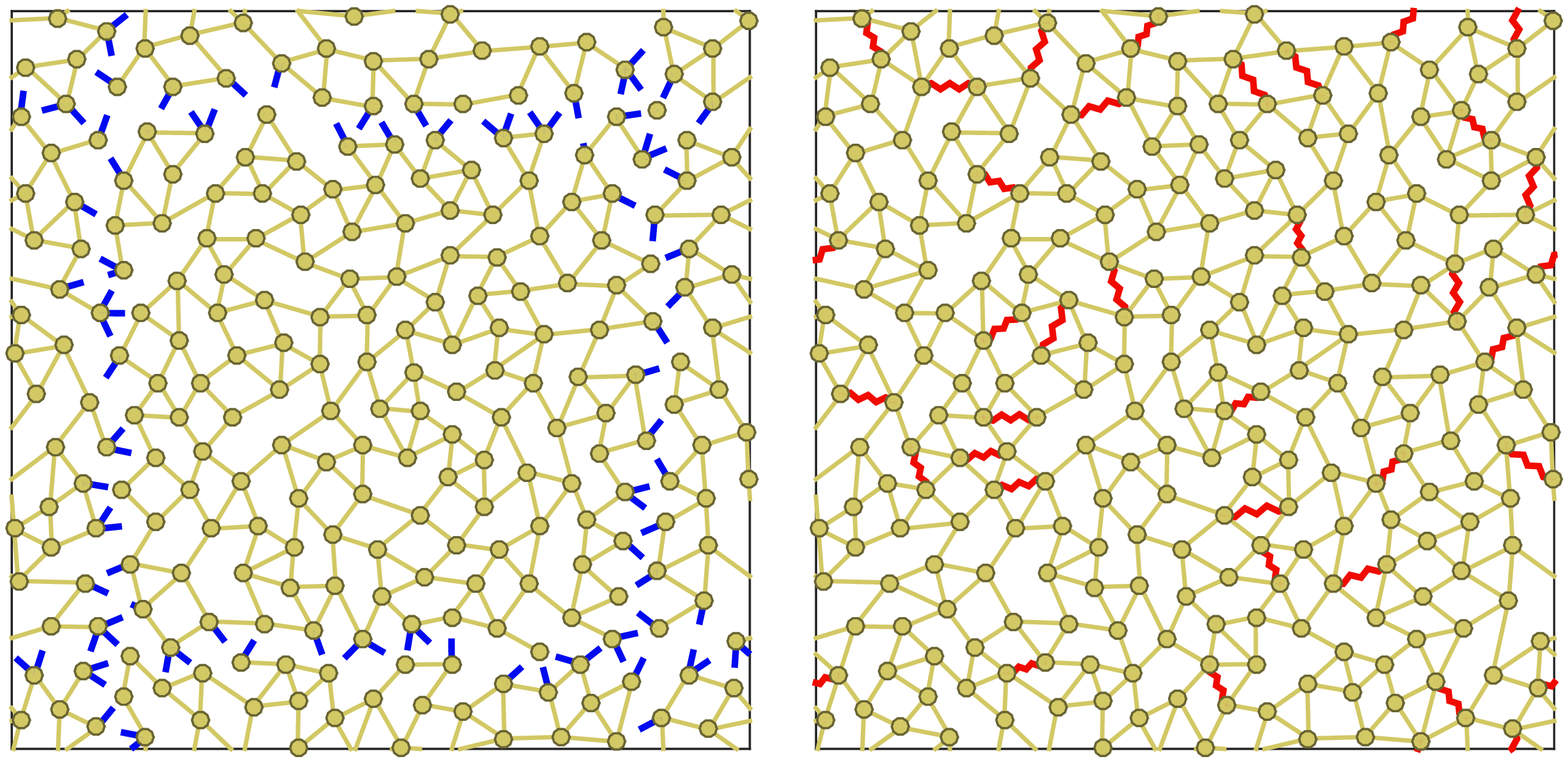}
\caption{ Left: illustrative diagram of the previous cutting argument of \cite{Wyart05}, showing cut bonds in blue (dashed lines). Bonds are cut around blocks of size $L\times L$. Right: in the new variational argument, bonds are now {\it added} as represented in red (broken lines). They are used to generate dipolar responses (by elongating the extra bonds individually), whose energy in the original system can be controlled. \label{f2}}
\end{figure}

Packings of soft particles are a very convenient system to study these soft elastic modes, as vibrational properties display scaling near their jamming transition \cite{Hecke10,Liu10,Ohern03}. In particular, there is a frequency scale $\omega^*\sim \delta z=z-z_c$  beyond which the density of vibrational modes $D(\omega)$ displays a plateau 
\cite{Wyart05a,Silbert05}. Here  $z$ is the coordination, and $z_c=2d$ is the  isostatic value where $d$ is the spatial dimension. The transverse structure factor at $\omega^*$ indicates a correlation length $l_c\sim 1/\sqrt{\delta z}$  \cite{Silbert05}. These results can be derived using effective medium theory \cite{Wyart10a}.
However, the first proposed derivation for the density of states was variational \cite{Wyart05}. It used  Maxwell's result that in networks with $z<z_c$, floppy modes with no restoring force must appear. In  \cite{Wyart05}  planes of contacts as depicted in Fig.~\ref{f2} were cut to generate floppy modes, from which trial modes could be made.  Using the hypothesis that floppy modes extend through the system led to the correct scaling for $\omega^*$ and its associated plateau.   However, a very recent numerical work supports that most of the floppy modes are in fact localized near the boundaries where the system is cut \cite{Sussman15}, raising doubts on the applicability of that argument. Here we provide a novel variational argument that does not make any hypothesis on the geometry of floppy modes, and numerically confirm its validity. Interestingly, the argument requires to add springs instead of cutting them. The argument emphasizes that the characteristic volume of soft elastic modes scales as $1/\delta z$, which is also the volume where finite size effects set in.

%

\section{Notation and previous results}
We consider a network of $N$ nodes of unit mass in $d$ dimensions, connected by $N_{\rm s}\equiv zN/2$ unstretched harmonic springs of unit stiffness,  with periodic boundary conditions. At leading order, the change of elastic energy $\delta E$ from equilibrium is quadratic in a small displacement field $|{\bf\delta R}\rangle$: 
\be
\label{1}
\delta E=\frac{1}{2}\langle{\bf\delta R}|{\cal M}|{\bf\delta R}\rangle,
\ee
where we define the stiffness matrix $\mm$, whose eigenvalues are denoted $\omega^2$. The $\omega$'s are the frequencies of vibrational modes. $\mm$  can be written as $\mm={\cal S}^t{\cal S}$ where ${\cal S}$ (sometimes denoted ${\cal Q}^t$) relates the displacement field ${\bf \delta R} $ of nodes, of dimension $Nd$ to an extension field of the springs ${\bf \delta r}$, of dimension $N_{\rm s}$: ${\cal S}{\bf \delta R}={\bf \delta r}$. Our goal is to build a variational argument setting a lower bound on the density of frequencies, $D(\omega)$. 

To do so, we first recall a formalism previously used  to derive the elastic moduli near jamming \cite{Wyart05b}. Consider changing the rest lengths of the springs by some small amount ${\bf y}$. Letting the nodes relax leaves a residual energy $E$:
\be
\label{3}
E=\frac{1}{2}\min_{{\bf \delta R} }\langle {\cal S}{\bf \delta R}-{\bf y} |{\cal S}{\bf \delta R}-{\bf y}\rangle,
\ee
Eq.(\ref{3})  is equivalent to the statement that the energy is the squared-norm of the projection of ${\bf y}$ on the kernel of ${\cal S}^t$. 
We introduce an orthonormal basis  $\{{\bf f_p}\}$ for that kernel. The $\{{\bf f_p}\}$ are the modes of self-stress, i.e. the sets of contact forces that balance force on each node \cite{Calladine78}. If the initial system has no floppy modes, this kernel must be of dimension $\delta z N/2=N_{\rm s}-dN + {\cal O}(1)$.  Here the ${\cal O}(1)$ term accounts for global translational and rotational modes, which depends on the choice of boundary conditions. 
Eq.(\ref{3}) implies that:
\be
\label{4}
E=\frac{1}{2}\sum_{p=1}^{\delta z N/2} \langle {\bf f_p}|{\bf y}\rangle^2.
\ee
Changing the rest length of only one spring $\alpha$ by amplitude $\epsilon$ exerts a force dipole whose energy must follow:
\be
\label{4}
E=\frac{\epsilon^2}{2}\sum_{p=1}^{\delta z N/2}  f_{\alpha,p}^2.
\ee
Since the ${\bf f_p}$ are normalised, $\langle f_{\alpha,p}^2\rangle =1/N_{\rm s}$, where the angular bracket indicates averaging over all springs. Thus, the average response energy for stretching one spring is:
\be
\label{5}
E=\frac{\delta z}{2z}\epsilon^2,
\ee
which defines an average local elastic modulus  $G_l=\delta z/z$.
This formula is exact for any value of positive $\delta z$, until the system is isostatic-plus-one-contact, where we get: $G_l=1/(dN)$. 
It implies that scaling holds down to a finite size volume $N_{\rm f.s.}\sim 1/\delta z$, where there is of order one additional contact with respect to isostaticity, as confirmed numerically~\cite{Goodrich12}.  In what follows, we use this result to extend arguments obtained for systems with $\delta z \sim 1/N$ to arbitrary $\delta z > 1/N$.

Finally, from this formalism  it is easy to show that changing the rest length of some contact $\alpha$ in an isostatic-plus-one-contact system leads to an energy proportional to $\propto (f_{\alpha,1} f_{\beta,1})^2$  in any given spring $\beta$. Assuming that the $\{{\bf f_p}\}$ are of similar magnitude throughout the system then implies that the  response to a dipole spreads everywhere and does not decay with distance. 
This assumption must be true in packings, since  the $\{{\bf f_p}\}$ are proportional to the real physical forces  at the jamming point, which must carry a positive pressure everywhere.  This reasoning was confirmed numerically,  as one finds that stretching a bond $\alpha$ by some amount $\epsilon$ leads to a displacement of  all particles of order $\epsilon$ \cite{Lerner13a}.

\section{Variational argument} We seek to show that an isostatic system has  at least a constant density of vibrational modes. First, we add one new spring to the system, which we call  $\alpha$, initially at rest. Next, we change its rest length by some $\epsilon$. As stated above, the corresponding response $\delta {\bf R}_\alpha$ is extended, so its norm must follow:
\be
\label{007}
||\delta {\bf R}_\alpha||^2\sim N \epsilon^2.
\ee
 Likewise  on average the response energy $E_\alpha$ is of order $ E_\alpha\sim \epsilon^2/N$ as follows from Eq.(\ref{5}). $\mm$ is symmetric and positive definite. Using $\delta {\bf R}_\alpha$ as a trial mode for the stiffness matrix of the original isostatic system, we can bound its lowest frequency:
\be
\label{8}
\omega_{\rm min}^2\leq \frac{2 E_\alpha}{||\delta {\bf R}_\alpha||^2}\sim \frac{1}{N^2}.
\ee

\begin{figure*}[!ht]
\centering
\includegraphics[width=1.8\columnwidth]{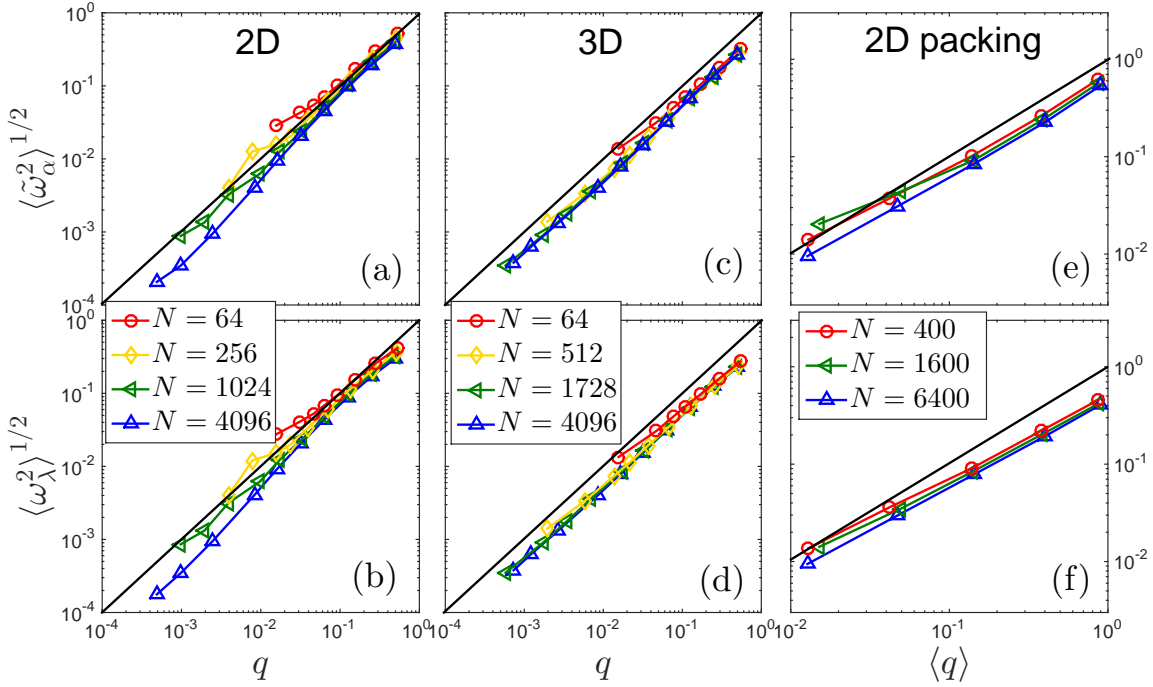}
\caption{ Average characteristic frequency $\langle \tilde \omega_\alpha^2\rangle^{1/2}$ of the response to spring elongations $\delta {\bf R}_\alpha$ defined in Eq.(\ref{10}) {\it vs} fraction of additional springs $q$ for 2D random networks (a), 3D random networks (c) and 2D packings (e). 
The average is made on the $Nq$ added springs. 
Bottom (b), (d) and (f): Average characteristic frequency ${\langle \omega_\lambda^2\rangle }^{1/2}$ of the orthonormal trial modes in the isostatic system. Black lines show the scaling of the bound $\omega=q$. 
\label{f3}}
\end{figure*}
Next, we extend this argument to finite $\omega$ by introducing a density $q$ of additional springs to an isostatic system, as sketched in Fig.~\ref{f2}.  To ensure a homogeneous distribution, we decompose the system into blocks of  volume  $N_{\rm f.s.}=1/q$,  that each contains  one additional spring with respect to the initial isostatic state.  The obtained system is hyperstatic with an excess coordination $\delta z=2q$. We denote by $\tilde E$ and $E$ the energy of a given displacement  in the hyperstatic and isostatic system, respectively. Since the isostatic system has fewer springs,  $\tilde E>E$. We label the introduced springs $\alpha$, with $\alpha=1, ..., qN$. We consider  the response $\delta {\bf R}_\alpha$ obtained in the hyperstatic system by changing the rest length of one spring $\alpha$ by some amount $\epsilon$. According to Eq.(\ref{5}), the energy of such a mode is ${\tilde E}_\alpha\sim \epsilon^2 q$. The norm of the response must scale as  $||\delta {\bf R}_\alpha||^2\sim \epsilon^2 q^{\gamma}$ where $\gamma$ is some exponent. Requiring that this expression matches Eq.\ref{007} when finite-size effects set in, i.e. $N=1/\delta z$, implies $\gamma=-1$ or  $||\delta {\bf R}_\alpha||^2\sim \epsilon^2/q$.  Defining
\be
\label{10}
{\tilde \omega}_{\alpha}\equiv \frac{\sqrt{2{\tilde E}_{\alpha}}}{||{\bf\delta R_{\alpha}}||},
\ee
we must then have $\tilde{\omega}_\alpha\sim q$, a scaling we numerically confirm in Fig.~\ref{f3}. Since $\omega_\alpha<\tilde{\omega}_\alpha$, if the vectors $\delta {\bf R}_\alpha$  for different $\alpha$ were orthogonal, we could use them as trial modes in Horn's variational theorem \cite{Wyart05} and obtain that there are at least of order $Nq/2$ normal  modes of frequency of order $\omega_{\alpha}$ or smaller in the isostatic system. As we discuss below, orthogonality is not a problem. It can  be anticipated  by writing  the norm of the response $\delta {\bf R}_\alpha$  in terms of a characteristic volume $V_c$ and the characteristic displacement $\epsilon$: $||\delta {\bf R}_\alpha||^2\equiv V_c \epsilon^2$. Comparing this expression with the results above, we get  $V_c\sim 1/q$. Since there  is a density $q$ of such responses, that each occupy a volume $1/q$, we only expect a limited overlap between these modes.
Assuming this to be true,  we have found $Nq$ modes with frequency smaller than $\omega$. Hence the number of modes per unit volume with frequency smaller than $\omega$, ${\cal N}(\omega)$, satisfies ${\cal N}(\omega) \geq Nq/N \sim \omega$, since here $\omega\sim q$. If this bound is saturated, one gets for the density of states:   
\be
\label{11}
D(\omega) = d{\cal N}/d\omega\sim  \omega^0
\ee
as observed in packings. 

This argument is readily extended to hyperstatic systems with  $\delta z>1/N$. Adding a density $q$ of additional springs, we create a system of excess coordination $\Delta z = \delta z + 2q.$ The only difference from the isostatic case is that the frequency of trial modes now reads $\omega_{\alpha}\sim \Delta z \sim q +\delta z$. Since trial modes all get a frequency shift $\delta z$ with respect to the isostatic network, our argument predicts that the density of states is simply shifted from that of an isostatic system, by a frequency  $\omega^*\sim \delta z$ below which trial modes cannot be produced, in agreement with numerical measurements of $D(\omega)$ \cite{Wyart05a,Silbert05}.

\section{Orthogonality of the trial modes}
For the variational argument to work, we must build from the displacement fields $\{ \delta {\bf R}_\alpha \}$ of the order $qN$ orthogonal trial modes, whose frequency remains of order $\omega \sim q$. This is achieved  by considering the correlation matrix ${\cal C}$ of dimension $qN\times qN$:
\be
\label{12}
{\cal C}_{\alpha,\beta}=\frac{\langle \delta {\bf R}_\alpha|\delta {\bf R}_\beta\rangle}{||\delta {\bf R}_\alpha|| \cdot ||\delta {\bf R}_\beta||}
\ee
${\cal C}$ is symmetric and can be diagonalised in an orthonormal basis $\delta {\bf R}_\lambda$:
\be
\label{13}
{\cal C}=\sum_\lambda \lambda |\delta {\bf R}_\lambda\rangle\langle  \delta {\bf R}_\lambda|
\ee
The $\delta {\bf R}_\lambda$ are our trial modes. We define, using the stiffness matrix of the isostatic system:
\be
\label{13} 
\omega_\lambda^2= \langle  \delta {\bf R}_\lambda|\mm |\delta {\bf R}_\lambda\rangle
\ee
A sufficient condition for our results to hold true is:
\be
\label{14}
\langle \omega_\lambda^2\rangle <C_0 q^2,
\ee
 where $C_0$ is some numerical constant  independent of $N$ and the average is made over the $qN$ modes $\delta {\bf R}_\lambda$. Indeed, in that case, at least $qN/2$ modes satisfy $\omega_\lambda<{\sqrt C_0} q$. Fig.\ref{f3} confirms that this condition holds true. 
 
%

\section{Numerical evidence}  
We consider two types of  networks:
({\it i}) An isostatic network can be built by removing redundant springs from a hyperstatic network. The latter is built from a strongly compressed bi-disperse soft particles with rest size ratio $1.4:1$, for which the forces in the bonds are set to zero. Next, we find and remove the spring connecting the nodes with the highest coordination, until isostaticity is reached.  The ``red bonds" shown in the right panel of Fig.~\ref{f2} are precisely  those that were removed to reach isostaticity; the first to be added are the last that were removed in that process. We consider both two dimension (2D) and three dimension (3D) with $N$ varying from $64$ to $4096$ and $q$ varying from $1/N$ to $0.5$. 
({\it ii}) The contact network of a  packing at jamming is also isostatic.
We construct such isostatic networks by decompressing the bi-disperse soft particle packings with the same size ratio as above. We track the configurations during the decompression. The  red springs we add in our arguments are precisely chosen as the contact lost when the packing was decompressed; once again, the first red springs we add are the last contacts which opened in the decompression protocol.  We consider two dimensional packings of size $N=400$ to $6400$ at pressure $p=1\times10^{-5}$ to $1\times10^{-1}$.

We compute frequencies as defined in Eq.(\ref{10}) both for the response to elongating one spring $|{\bf\delta R_{\alpha}}\rangle$ and the orthogonal trial modes $|{\bf\delta R_{\lambda}}\rangle$. This is done for the two types networks as $q$ is varied. Average values of for these frequencies are shown in Fig.~\ref{f3}. 
Both our scaling prediction $\omega_\alpha \sim q$ and  Eq.(\ref{14}) are found to be satisfied. The inequality Eq.(\ref{14}) becomes better satisfied with growing system size, indicating that it holds in the thermodynamic limit, and proving the validity of our argument.



\section{Discussion} 
We have argued that  a variational argument for packings near jamming can be made by exciting the system in its bulk, rather than at its boundary. The physical picture it provides is that soft elastic modes of frequency $\omega$ at jamming are similar to the response to a local strain in a system of coordination $\delta z\sim \omega$.  Away from jamming (i.e. for $\delta z>0$),  the softest elastic modes can be localized (in the sense of fixing the ratio of their squared norm over their squared maximum displacement)  on a volume $1/\omega^*\sim 1/\delta z$ without significantly increasing their frequency. This picture   shares similarity with the  description of dense granular and suspensions flows presented in  \cite{Degiuli15a}. In that case, the contact network is hypostatic ($\delta z<0$), and flow occurs along floppy modes whose characteristic volume is precisely $\sim 1/|\delta z|$, which also corresponds to the characteristic finite size volume beyond which scaling holds. 

The characteristic volume $N_{\rm f.s.}\sim 1/\delta z$ may seem to contradict effective medium results and numerics, which indicates that the  length scale  characterizing the decay of most of the amplitude of the response to a local strain is $l_c\sim 1/\sqrt{\delta z}$ both for floppy \cite{During13} and jammed materials \cite{Lerner14}. There is however no contradiction:  it simply signals that the leading term of the response  decays  with distance $r$ as $\delta R^2_\alpha(r)\sim f(r/l_c)/r^{d-2}$, where $f(x)$ is a rapidly decaying function of its argument.

We expect  our argument to be robust to disorder in the spring stiffness, as long as they are not too many weak springs in the system.
When this occurs, better variational arguments can be designed that cut the weakest springs and consider the resulting floppy modes as trial modes.
These arguments are necessary to describe thermal effects on vibrational properties in packings of soft particles \cite{Degiuli15} as well as hard sphere systems \cite{DeGiuli14b}.


Finally, we expect our lower bound on the density of states to hold for generic isostatic networks. An interesting case are ``directed" isostatic networks  where the geometry of floppy modes can be computed layer by layer. These networks can present  interesting topological properties \cite{Kane14,Paulose15,Lubensky15}. They also present curious finite size effects \cite{Moukarzel12} when mixed  boundary conditions (one side pinned and one side free) are used.  We expect however that finite size effects are similar to those discussed here when periodic boundaries are used in disordered networks. We have checked that our prediction on the linear dependence of $\omega_\alpha$ and $\omega_\lambda$ with the density of added springs $q$ holds true for a distorted square lattice, which is ``directed".

\acknowledgements
We are grateful to T.~Lubensky and D.~Sussman for discussions and for sharing unpublished data, and thank E. Lerner for discussions and providing packings, and  G. ~D\"uring, J.~Lin, and T. Witten for discussions. LY was supported in part by the National Science Foundation under Grant No. NSF PHY11-25915. This material is based upon work performed using computational resources supported by the ``Center for Scientific Computing at UCSB'' and NSF Grant CNS-0960316.

\bibliographystyle{ieeetr}
\bibliography{../bib/Wyartbibnew}

\begin{thebibliography}{10}

\bibitem{Phillips81}
A.~Anderson, {\em Amorphous Solids: Low Temperature Properties}, vol.~24 of
  {\em Topics in Current Physics}.
\newblock Springer, Berlin, 1981.

\bibitem{Baldi10}
G.~Baldi, V.~Giordano, G.~Monaco, and B.~Ruta, ``Sound attenuation at terahertz
  frequencies and the boson peak of vitreous silica,'' {\em Physical Review
  Letters}, vol.~104, no.~19, p.~195501, 2010.

\bibitem{Xu09}
N.~Xu, V.~Vitelli, M.~Wyart, A.~J. Liu, and S.~R. Nagel, ``Energy transport in
  jammed sphere packings,'' {\em Physical Review Letters}, vol.~102,
  pp.~038001--, 01 2009.

\bibitem{Leonforte06}
F.~Leonforte, A.~Tanguy, J.~Wittmer, and J.-L. Barrat, ``Inhomogeneous elastic
  response of silica glass,'' {\em Physical review letters}, vol.~97, no.~5,
  p.~055501, 2006.

\bibitem{Lerner14}
E.~Lerner, E.~DeGiuli, G.~D{\"u}ring, and M.~Wyart, ``Breakdown of continuum
  elasticity in amorphous solids,'' {\em Soft Matter}, vol.~10, pp.~5085--5092,
  2014.

\bibitem{Karimi15}
K.~Karimi and C.~E. Maloney, ``Elasticity of frictionless particles near
  jamming,'' {\em Physical Review E}, vol.~92, no.~2, p.~022208, 2015.

\bibitem{Ellenbroek06}
W.~G. Ellenbroek, E.~Somfai, M.~van Hecke, and W.~van Saarloos, ``Critical
  scaling in linear response of frictionless granular packings near jamming,''
  {\em Phys. Rev. Lett.}, vol.~97, p.~258001, Dec 2006.

\bibitem{Brito07}
C.~Brito and M.~Wyart, ``Heterogeneous dynamics, marginal stability and soft
  modes in hard sphere glasses,'' {\em Journal of Statistical Mechanics: Theory
  and Experiment}, vol.~2007, no.~08, p.~L08003, 2007.

\bibitem{Hocky12}
G.~M. Hocky and D.~R. Reichman, ``A small subset of normal modes mimics the
  properties of dynamical heterogeneity in a model supercooled liquid,'' {\em
  The Journal of chemical physics}, vol.~138, no.~12, p.~12A537, 2013.

\bibitem{Manning11}
M.~L. Manning and A.~J. Liu, ``Vibrational modes identify soft spots in a
  sheared disordered packing,'' {\em Phys. Rev. Lett.}, vol.~107, p.~108302,
  Aug 2011.

\bibitem{Wyart05a}
M.~Wyart, L.~E. Silbert, S.~R. Nagel, and T.~A. Witten, ``Effects of
  compression on the vibrational modes of marginally jammed solids,'' {\em
  Physical Review E}, vol.~72, no.~5, p.~051306, 2005.

\bibitem{Brito09}
C.~Brito and M.~Wyart, ``Geometric interpretation of previtrification in hard
  sphere liquids,'' {\em The Journal of Chemical Physics}, vol.~131, no.~2,
  pp.~024504--024518, 2009.

\bibitem{DeGiuli14b}
E.~DeGiuli, E.~Lerner, C.~Brito, and M.~Wyart, ``Force distribution affects
  vibrational properties in hard-sphere glasses,'' {\em Proceedings of the
  National Academy of Sciences}, vol.~111, no.~48, pp.~17054--17059, 2014.

\bibitem{Florijn14}
B.~Florijn, C.~Coulais, and M.~van Hecke, ``Programmable mechanical
  metamaterials,'' {\em Physical review letters}, vol.~113, no.~17, p.~175503,
  2014.

\bibitem{Kane14}
C.~Kane and T.~Lubensky, ``Topological boundary modes in isostatic lattices,''
  {\em Nature Physics}, vol.~10, no.~1, pp.~39--45, 2014.

\bibitem{Paulose15}
J.~Paulose, B.~G.-g. Chen, and V.~Vitelli, ``Topological modes bound to
  dislocations in mechanical metamaterials,'' {\em Nature Physics}, 2015.

\bibitem{Hecke10}
M.~van Hecke, ``Jamming of soft particles: geometry, mechanics, scaling and
  isostaticity,'' {\em Journal of Physics: Condensed Matter}, vol.~22, no.~3,
  pp.~033101--033124, 2010.

\bibitem{Liu10}
A.~J. Liu, S.~R. Nagel, W.~van Saarloos, and M.~Wyart, {\em The jamming
  scenario: an introduction and outlook}.
\newblock Oxford: Oxford University Press, 2010.

\bibitem{Ohern03}
C.~S. O'Hern, L.~E. Silbert, A.~J. Liu, and S.~R. Nagel, ``Jamming at zero
  temperature and zero applied stress: The epitome of disorder,'' {\em Phys.
  Rev. E}, vol.~68, pp.~011306--011324, Jul 2003.

\bibitem{Silbert05}
L.~E. Silbert, A.~J. Liu, and S.~R. Nagel, ``Vibrations and diverging length
  scales near the unjamming transition,'' {\em Phys.\ Rev.\ Lett.}, vol.~95,
  p.~098301, 2005.

\bibitem{Wyart10a}
M.~Wyart, ``Scaling of phononic transport with connectivity in amorphous
  solids,'' {\em EPL (Europhysics Letters)}, vol.~89, no.~6, p.~64001, 2010.

\bibitem{Wyart05}
M.~Wyart, S.~Nagel, and T.~Witten, ``Geometric origin of excess low-frequency
  vibrational modes in weakly connected amorphous solids,'' {\em EPL
  (Europhysics Letters)}, vol.~72, no.~3, pp.~486--492, 2005.

\bibitem{Sussman15}
D.~M. Sussman, O.~Stenull, and T.~Lubensky, ``Topological boundary modes in
  jammed matter,'' {\em arXiv preprint arXiv:1512.04480}, 2015.

\bibitem{Wyart05b}
M.~Wyart, ``On the rigidity of amorphous solids,'' {\em Annales de Phys},
  vol.~30, no.~3, pp.~1--113, 2005.

\bibitem{Calladine78}
C.~Calladine, ``Buckminster fuller's ``tensegrity'' structures and clerk
  maxwell's rules for the construction of stiff frames,'' {\em International
  Journal of Solids and Structures}, vol.~14, no.~2, pp.~161 -- 172, 1978.

\bibitem{Goodrich12}
C.~P. Goodrich, A.~J. Liu, and S.~R. Nagel, ``Finite-size scaling at the
  jamming transition,'' {\em Physical review letters}, vol.~109, no.~9,
  p.~095704, 2012.

\bibitem{Lerner13a}
E.~Lerner, G.~During, and M.~Wyart, ``Low-energy non-linear excitations in
  sphere packings,'' {\em Soft Matter}, vol.~9, pp.~8252--8263, 2013.

\bibitem{Moukarzel12}
C.~F. Moukarzel, ``Elastic collapse in disordered isostatic networks,'' {\em
  EPL (Europhysics Letters)}, vol.~97, no.~3, p.~36008, 2012.

\bibitem{Degiuli15a}
E.~DeGiuli, G.~D\"uring, E.~Lerner, and M.~Wyart, ``Unified theory of inertial
  granular flows and non-brownian suspensions,'' {\em Physical Review E},
  vol.~91, p.~062206, 06 2015.

\bibitem{During13}
G.~D{\"u}ring, E.~Lerner, and M.~Wyart, ``Phonon gap and localization lengths
  in floppy materials,'' {\em Soft Matter}, vol.~9, no.~1, pp.~146--154, 2013.

\bibitem{Degiuli15}
E.~DeGiuli, E.~Lerner, and M.~Wyart, ``Theory of the jamming transition at
  finite temperature,'' {\em The Journal of chemical physics}, vol.~142,
  no.~16, p.~164503, 2015.

\bibitem{Lubensky15}
T.~Lubensky, C.~Kane, X.~Mao, A.~Souslov, and K.~Sun, ``Phonons and elasticity
  in critically coordinated lattices,'' {\em arXiv preprint arXiv:1503.01324},
  2015.

\end{thebibliography}

\end{document}